\documentclass[10pt, conference]{ieeeconf}

\IEEEoverridecommandlockouts                              
\usepackage{graphics} 
\usepackage{epsfig} 
\usepackage{mathptmx} 
\usepackage{times} 
\usepackage{amsmath} 
\usepackage{amssymb}  

\usepackage{psfrag}
\usepackage{subfigure}

\def\reals{\hbox{I\kern -.19em R}}
\newtheorem{proposition}{Proposition}
\newtheorem{theorem}{Theorem}
\newtheorem{definition}{Definition}
\newtheorem{corollary}{Corollary}
\newtheorem{lemma}{Lemma}
\newcommand{\myBox}{\hfill\rule{2mm}{2mm}}

\title{ \vspace{-10mm}\Huge Dynamical Structure Functions for the Estimation of LTI Networks with Limited Information\vspace{.5cm}}
\author{\hspace{.75in}\begin{tabular}{c}Jorge Gon\c{c}alves\\Control Group\\Department of Engineering\\Cambridge University\\Cambridge CB2 1PZ, UK\\{\tt jmg77@cam.ac.uk}\end{tabular}\hspace{1in}\begin{tabular}{c}Sean Warnick\\Information and Decision Algorithm Laboratories\\Department of Computer Science\\Brigham Young University\\Provo, UT 84602, USA\\{\tt sean@cs.byu.edu}\end{tabular}}

\begin{document}
\bibliographystyle{plain}

\maketitle

\vspace{-1.65cm}
\thispagestyle{empty}

\begin{abstract}
  This research explores the role and representation of network
  structure for LTI Systems.  We demonstrate that transfer functions
  contain no structural information without more assumptions being
  made about the system, assumptions that we believe are unreasonable
  when dealing with truly complex systems.  We then introduce
  Dynamical Structure Functions as an alternative, graphical-model
  based representation of LTI systems that contain both dynamical and
  structural information of the system.  We use Dynamical Structure to
  prove necessary and sufficient conditions for estimating structure
  from data, and demonstrate, for example, the danger of attempting to
  use steady-state information to estimate network structure.
  
 \end{abstract}

\section{Introduction}

One of the fundamental issues for modeling, identifying, and
controlling complex networked systems is inferring system structure
from input-output data.  Structure is often the key for understanding
a variety of complex systems because it enables a decomposition of the
complete system into an interconnection of subsystems.  When analysis
of the subsystems is comparatively simple, and the interconnection
structure is well understood, then the behavior of the complex system
can be deduced from an understanding of its components.  Moreover,
exploiting structural information can tremendously reduce the
conservatism of robust solutions designed to compensate for system
uncertainty.  This impact on complexity and uncertainty makes
structural information extremely important in the analysis of complex
networked systems.

Examples of scientists working on identifying or exploiting network
structure arise in a variety of disciplines.  Social scientists have
developed a rich literature on the use of network models to describe
interpersonal associations, perhaps one of the most famous works being
Milgram's "small world" experiment in the 1960's in which letters
passed from person to person were able to reach a particular target
individual in only about six steps \cite{Milgram}.  More recently,
attention has focused on networks of business communities
\cite{Galaskiewicz,Mariolis, Mizruchi}, internet-enabled virtual
communities \cite{Holme}, citation networks in scientific communities
\cite{Redner, Seglen}, preference networks for product recommender
systems \cite{Goldberg, Resnick}, distribution networks \cite{Amaral},
and the detection and destabilization of terrorist networks
\cite{carley}.  Epidemiologists have developed models for the dynamics
of both epidemic and endemic diseases spreading through population
networks \cite{Hethcote}, computer scientists have developed
algorithms for searching over networks that are deployed in a number
of popular applications \cite{Brin}, and biologists use microarray and
other data sources to infer the regulation structure in genomic,
proteomic, and metabolic networks \cite{Guelzim, Maslov, Stelling,
  Uetz} .

Discovering structure from data, however, can be difficult.  Typical
identification methods do not emphasize structure estimation, but
rather focus on behavior generalization by selecting a model that
accurately predicts system outputs for unobserved inputs.  As long as
the dynamic behavior of the system is accurately described, the
question of structure is often avoided altogether.  For many
applications, various model structures for the same input-output map
are equally useful for forecasting and control.  Nevertheless,
sometimes it is important not only to describe the system dynamics
accurately, but to do so with a model that correctly represents the
structure of the original system.

In contrast with these identification methods that emphasize system
dynamics over structure, inference methods have been developed that
emphasize structure over dynamics.  These methods employ graphical
models to describe network structure.  Nodes represent system states,
understood to be random variables, and edges indicate conditional
dependence between variables.  Using Bayes rule, measurements can then
be used to update prior distributions believed to characterize
relationships throughout the network.  A rich literature has grown in
this area, and even issues of inferring causality from correlation
have been addressed at some level \cite{Jensen, Jordan, Pearl}.

Nevertheless, although these Bayesian Networks provide an efficient
way to parameterize the joint probability distribution characterizing
the entire system, conditional probabilities do not capture system
dynamics, and the most successful inference techniques only work on
directed acyclic graphs \cite{Cowell}.  For some applications, such as
modeling the citation network for a particular body of research,
assuming the network is acyclic is reasonable since papers generally
only cite previously published work.  There are many applications,
however, such as modeling biological or social or economic networks,
where such an assumption insisting on the absence of feedback
dependencies between system states would be entirely unreasonable.
Moreover, often an accurate representation of system dynamics is as
important as that of system structure.  In these situations, new
methods are needed.

This paper introduces Dynamical Structure Functions as a structurally
accurate representation of complex LTI systems that do not ignore
system dynamics.  We begin in the next section by demonstrating that
transfer functions contain no structural information without more
assumptions being made about the system, assumptions that we believe
are unreasonable when dealing with truly complex systems.  We also
highlight some common pitfalls when estimating structure from data.
We then introduce in Section \ref{se:net} the Dynamical Structure
Function of an LTI system and discuss its properties.
Section~\ref{se:main} then uses Dynamical Structure to provide
necessary and sufficient conditions for estimating structure from
data, and an example is provided illustrating the danger of using only
steady-state information to estimate structure. Section~\ref{se:conc}
then concludes with a discussion of future work.

\section{Background: Structure Estimation and Dynamic Systems}\label{se:mpf}

Consider the network characterized by the linear system
\begin{equation}\label{eq:LTI}
  \left\{\begin{array}{lll}
      \dot x & = & Ax + Bu \\
      y & = & Cx
    \end{array}\right .
\end{equation}
where $x\in R^n$, $u\in R^m$, $y\in R^p$, and $C=[I \ \ 0]$.  We are
interested in inferring the causal dependencies between the $p$
measured states, $y$, from limited data.  Typically, $m<n$, $p<n$, and $n$
itself is unknown. 

In this work we do not assume that the system (\ref{eq:LTI}) is both controllable and observable from the particular inputs and outputs specified by $u$ and $y$.  In the complex systems context, such an assumption would be unreasonable to impose since the number of inputs and outputs is assumed to be very small compared to the total number of states.  Thus, assuming controllability and observability would be restricting our attention to networks with very special structure.  As a result, we can not assume that (\ref{eq:LTI}) is a minimal realization of the corresponding input-output transfer
function, $G$, given by
\begin{equation}\label{eq:G}
G(s) = C\left ( sI -A \right ) ^{-1}B
\end{equation}

In this work we also do not assume knowledge of the system's order.  Thus, the true system, (\ref{eq:LTI}), has a particular causal structure and complexity that we can only detect through our interaction with the system at $u$ and $y$.  Nevertheless, we do assume
throughout this work that the transfer function, $G$, can be obtained
from the available data, $u$ and $y$, using standard identification
methods.

Notice that the transfer function does not directly reveal structural
information of the system.  For example, consider the system
\begin{equation}
\label{eq:TFexample}
A=\left[\begin{array}{cccc}-1&0&0&1\\ .25&-1&0&0\\
    0&1&-1&0\\0&0&.25&-1\end{array}\right]\ \ \ \ \ 
B=\left[\begin{array}{ccc}1&0&0\\0&1&0\\0&0&1\\0&0&0\end{array}\right]
\end{equation}
\[
C=\left[\begin{array}{cccc}1&0&0&0\\0&1&0&0\\0&0&1&0\end{array}\right]
\]
Note that this system has a very definite ring structure, where $x_1\longrightarrow x_2 \longrightarrow x_3 \longrightarrow x_4 \longrightarrow x_1$.  Nevertheless, the associated transfer function, $G$, is given by
\[\small
 \left[\begin{array}{ccc} s^3+3s^2+3s+1 &.25 & .25s+.25 \\ 
     .25s^2+.5s+.25 & s^3+3s^2+3s+1 & .625 \\ 
     .25s+.25 & s^2+2s+1 & s^3+3s^2+3s+1\end{array}\right] \frac 1 {p(s)}
\]
where $p(s)=s^4+4s^3+6s^2+4s+.9375$, which reveals nothing about the
ring structure of the system.  Although the structure is easy to read
from the actual state realization of the system, a transfer function
identified from input-output data--{\it even if identified
perfectly}--does not directly yield any structural information about
the system.

Given this difficulty using the transfer function to obtain structural
information, one may ask why not identify the state space realization
directly.  Nevertheless, it is difficult to identify a realization of
the system without knowing the order of the system.  In this work, we
assume that structural information must be obtained from limited data,
that is, with measurements that constitute only part of the complete
state vector.  Moreover, we do not assume knowledge of the full system complexity, or true system order.  Later, we demonstrate how incorrectly assuming knowledge of
the system order can lead to erroneous structural estimates.

Thus, transfer functions are generally obtainable from input-output data, but they contain no structural information.  At the other extreme, state space realizations contain all information about the system, but they are difficult to obtain from limited information.  We are interested in something in between, a representation that may still be obtainable from input-output data, but that also contains information about both the dynamics and the structure of the system.

Structure is typically represented by a graph.  Nodes represent system
variables, and edges represent interaction between variables.
Directed edges capture notions of directed influence, often quantified
by conditional probabilities.  We will consider a directed edge to
indicate a causal relationship between variables.  Regardless of how
the notion of directed influence is represented, however, the absence
of an edge between variables indicates a kind of independence between
those variables; $z_1 \longrightarrow z_2 $ instead of $z_1
\rightleftharpoons z_2$ means that $z_1$ does not depend directly on
$z_2$.  That is, any influence $z_2$ may have on $z_1$ may only occur
indirectly through $z_2$'s influence on other {\it explicit} variables
(nodes) in the network, and their direct influence, in turn, on $z_1$.
In particular, it is critical to note that $z_1 \rightarrow z_2$ means
there may {\it not} be some hidden variable, $z_i$, {\it that has not
  been represented in the graphical network model} through which $z_2$
influences $z_1$.  For structure to have meaning, even hidden,
unmodeled variables should respect the graph defining the network and
only operate within edges.  This has important implications for
dealing with uncertainty.

In its simplest form, then, structure is simply a square binary matrix
$S$ with $s_{ij}=1$ indicating the presence of an edge directed from
$z_j$ to $z_i$.  For the system $T$ given in (\ref{eq:LTI}) we would
define our explicitly modeled variables to be $z = [z1;\;z2] [y;\;u]$.  Simple structure, $S$, would then be a $p+m$ by $p+m$
binary matrix; for the example (\ref{eq:TFexample}) we would have:
\begin{equation}
\label{eq:structure}
S = \left[\begin{array}{cc}Q_T&P_T\\P_F&Q_F\end{array}\right]=\left[\begin{array}{ccc|ccc}1&1&0&1&0&0\\ 0&1&1&0&1&0\\ 1&0&1&0&0&1\\ \hline 0&0&0&1&0&0
\\0&0&0&0&1&0\\ 0&0&0&0&0&1\end{array}\right].
\end{equation}

Note that we consider that variables may automatically influence themselves since they may be recursively generated, thus the diagonal of $S$ is identity.  The blocks $Q_T$, $P_T$, $Q_F$, and $P_F$ correspond to the partition of $z$ as inputs and outputs of the system $T$.  The {\it input structure}, $P_T$ describes how inputs, $u$, influence the measured variables, $y$.  The {\it output structure}, $Q_T$ describes how the measured variables, $y$, influence each other.  Under the interpretation that $y$ corresponds to part of the system state vector, the output structure $Q_T$ may also be called the {\it internal structure} of the system $T$ (with $P_T$ then being called the {\it control structure} of $T$).  The remaining blocks, $Q_F$ and $P_F$, describe the feedback environment of $T$.  In general, when $T$ is in feedback with an operator $F$, $P_F$ is the input or control network of the feedback operator $F$, while $Q_F$ is $F$'s internal or output structure.  When no feedback operator is defined and the inputs $u$ are truly considered free variables, then $P_F$ is zero (since $u$ does not depend on $y$), and $Q_F$ is identity (since $u$'s only depend on themselves and do not influence each other). 

Just as a transfer function description of a system grows or shrinks with the number of inputs and outputs of the system, the structure matrix $S$ also grows or shrinks with the number of system inputs and outputs.  We call the number of outputs, $p$, the {\it $p^{th}$-order resolution} of the structural representation.  Thus, when three of the states of a fourth order system are measured, the resolution of the structural representation is three, and $Q_T$ will be $3\times 3$.  The fourth state is hidden and does not appear in the third-order resolution of $S$ in any way.  Nevertheless, correct structural representations are {\it consistent}, in that zeros appear in lower-order (coarser) resolutions only if there are no hidden states from higher-order (finer) resolutions that could enable the interaction.

Structure estimation for dynamic systems seeks to find $S$ corresponding to a particular realization of a dynamic system, $T$, using only input output data.  Before discussing how to solve this problem, however, we first outline two flawed approaches to this problem that appear from time-to-time in the literature.  First, one may assume knowledge of the system order, $n$, and then proceed to attempt to infer information about structure in light of this assumption.  Second, one may estimate a particular realization of $T$ and then attempt to reconstruct $S$ from the state space model.  These approaches are not entirely unrelated, but we show next that either approach can easily lead to incorrect conclusions.  

\subsection{Example: Erroneous System Order Assumption}\label{se:example1}

Although there are some reasonable techniques for estimating order from time-series data, there is no foolproof method available.  In some applications, the most common technique for order estimation continues to be to assume that the measured outputs constitute the entire state vector, that is, that $n=p$.  The following example demonstrates that making this assumption incorrectly may lead to completely erroneous structure estimates.

\begin{figure}[h] 
   \centering
  \subfigure[Network]{\label{fig:example1a}
    \psfrag{u}[rt][rt]{$u$}
    \psfrag{x1}[rt][rt]{$x_1$}
    \psfrag{x2}[rt][rt]{$x_2$}
    \psfrag{x3}[rt][rt]{$x_3$}
    \includegraphics[width=3cm]{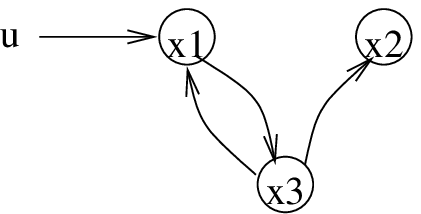}}
  \subfigure[Step response of $x_1$ and $x_2$]{\label{fig:example1b}
    \includegraphics[width=5cm]{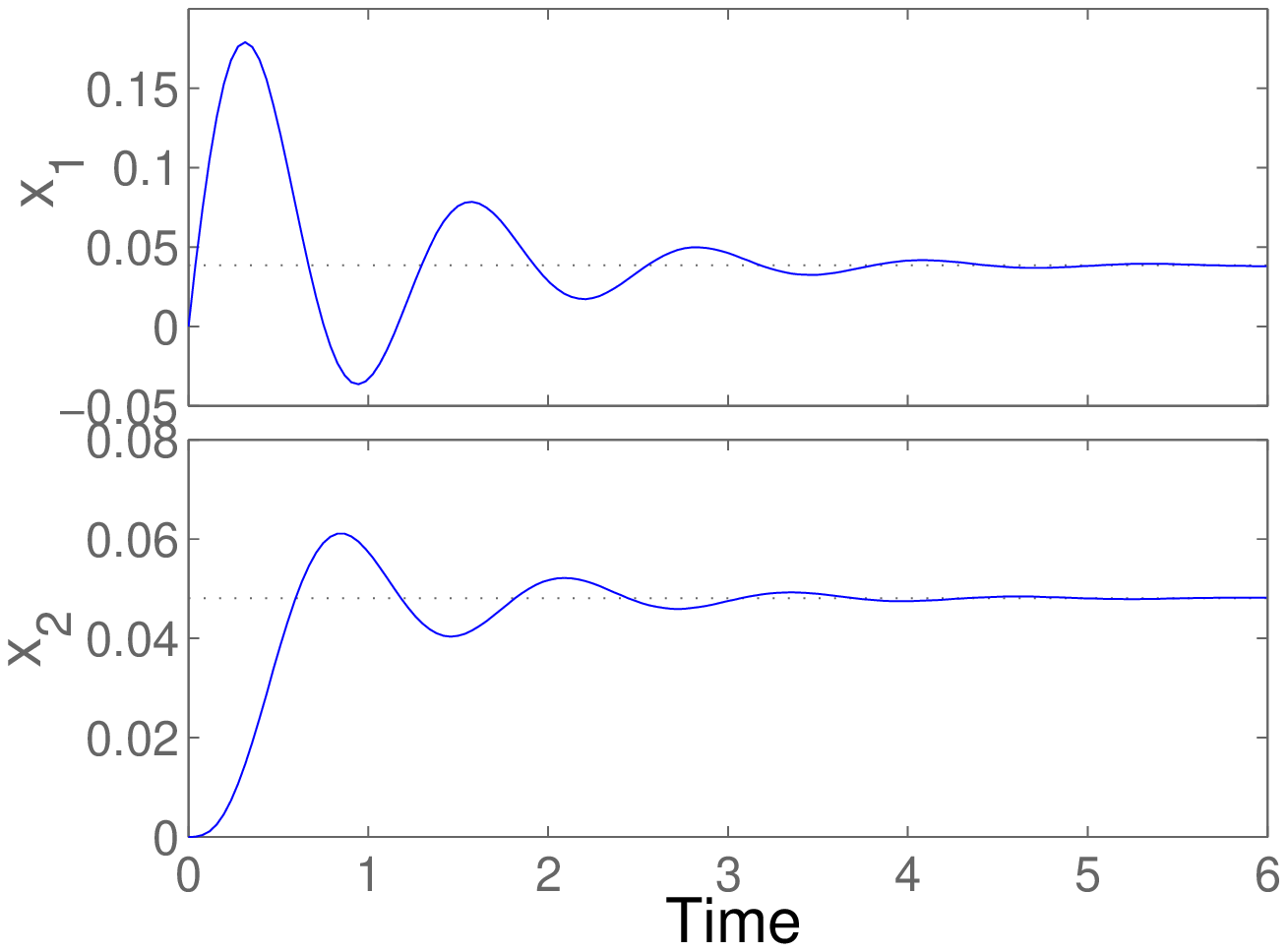}}
   \caption{Example of a simple 3 state network}\label{fig:example1}
\end{figure}

Consider the network in Figure~\ref{fig:example1a} with three state
variables structured in a chain, with the single input $u$ driving
$x_1$, $x_1$ in feedback with $x_3$, and $x_3$ driving $x_2$,
characterized by the equations
\begin{equation}\label{eq:example1}
  \left[\begin{array}{c}\dot{x}_1\\\dot{x}_2\\\dot{x}_3\end{array}\right] = 
  \left[\begin{array}{rrr}
      -1&0&-5\\ 0&-4&1\\ 5&0&-1
  \end{array}\right]\left[
  \begin{array}{c}x_1\\ x_2\\x_3\end{array}\right] +
  \left[\begin{array}{c}1\\0\\0\end{array}\right]u,
\end{equation}
$$
\left[\begin{array}{c}y_1\\ y_2\end{array}\right] \left[\begin{array}{rrr}1&0&0\\
    0&1&0\end{array}\right]\left[\begin{array}{c}x_1\\ x_2\\
    x_3\end{array}\right],$$
From $x_1$ and $x_2$, we would like to be able to infer the structure
$u\longrightarrow x_1 \longrightarrow x_2$, in spite of the fact that
we may have no knowledge of the not-directly-observed, yet
(indirectly) observable state $x_3$.

By assuming knowledge of the system order, one may then attempt to fit a state space realization
directly from the data.  In this case, any attempt to identify a second order system given the oscillating data shown above will result in an $A$ matrix with complex eigenvalues.  This implies that any real-valued $A$ matrix that reasonably fits the data will have non-zero terms in its off-diagonal positions, leading incorrectly to a fully connected network structure estimate instead of the correct chain structure.  
\myBox

\subsection{Example: Erroneous Structure from Realizations}\label{se:exam}

\begin{figure}[h]
  \centering \subfigure[Possible network 1.]{
    \label{fig:ex21a}
    \psfrag{u}[rt][rt]{$u$}
    \psfrag{x1}[rt][rt]{$y_1$}
    \psfrag{x2}[rt][rt]{$y_2$}
    \psfrag{x3}[rt][rt]{$x_3$}
    \includegraphics[width=3cm]{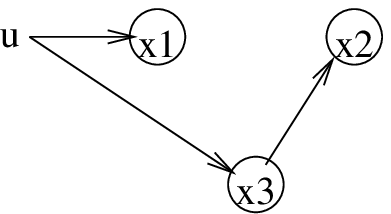}}
  \subfigure[Possible network 2.]{
    \label{fig:ex21b}
    \psfrag{u}[rt][rt]{$u$}
    \psfrag{x1}[rt][rt]{$y_1$}
    \psfrag{x2}[rt][rt]{$y_2$}
    \includegraphics[width=3cm]{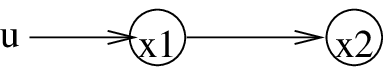}}
  \caption{Two possible networks given the data.}\label{fig:ex21}
\end{figure}

Suppose that after a sequence of experiments, one was able to identify
the transfer function
\begin{equation}\label{eq:examp_G}
G(s) \left[\begin{array}{c}\frac{1}{s+1}\\\frac{1}{(s+1)(s+2)}\end{array}\right].
\end{equation}
It can be shown that this transfer function is consistent with two
systems with very different structures, given by
$$A_1 = \left[\begin{array}{rrr}-1&0&0\\0&-2&1\\0&0&-1\end{array}\right]\ \ 
B_1=\left[\begin{array}{c}1\\0\\1\end{array}\right] \ \ 
C_1 = \left[\begin{array}{ccc}1&0&0\\0&1&0\end{array}\right]$$
and
$$A_2 = \left[\begin{array}{rr}-1&0\\1&-2\end{array}\right]\ \
B_2=\left[\begin{array}{c}1\\0\end{array}\right] \ \ C_2 \left[\begin{array}{cc}1&0\\0&1\end{array}\right]$$
The networks in Figure~\ref{fig:ex21} correspond to each of the
possible realizations of $G$.  Note that without more information
about the system, such that it is minimal, or order three, etc. then
we would not be able to say anything about structure from the transfer
function alone.

These examples demonstrate the difficulty of estimating network
structure from data.  Nevertheless, ideally one would estimate both
the network structure and the system dynamics from data. In the next
section, we introduce Dynamical Structure Functions as a mechanism for
representing both system dynamics and structure.


\section{Dynamical Structure}\label{se:net}

Consider the system given by (\ref{eq:LTI}).  Given the special structure on $C$, we note that the first p state variables are actually the measured variables $y$.  Defining $x_h$ to be the remaining $n-p$ ``hidden" states, the system becomes
\begin{equation}\label{eq:LTIhiddenstates}
  \left\{\begin{array}{lll}
      \left[\begin{array}{c}\dot{y}\\\dot{x}_h \end{array}\right]& = & \left[\begin{array}{cc}A_{11}&A_{12}\\A_{21}&A_{22}\end{array}\right]\left[\begin{array}{c}y\\x_h\end{array}\right] +\left[\begin{array}{c} B_{1}\\B_2\end{array}\right]u \\
      y & = & \left[\begin{array}{cc}I&0\end{array}\right]\left[\begin{array}{c}y\\x_h\end{array}\right]
    \end{array}\right .
\end{equation}
Taking Laplace Transforms of the signals, we then obtain
\begin{equation}\label{eq:LTIlaplace}
  \begin{array}{lll}
      \left[\begin{array}{c}sY\\sX_h \end{array}\right]& = & \left[\begin{array}{cc}A_{11}&A_{12}\\A_{21}&A_{22}\end{array}\right]\left[\begin{array}{c}Y\\X_h\end{array}\right] +\left[\begin{array}{c} B_{1}\\B_2\end{array}\right]U
      \end{array}
\end{equation}

From this equation it is easy to construct the transfer functions from the manifest variables $z = [Y\;;\;U]$ to themselves.  Solving for $X_h$, we have
$$X_h=\left ( sI - A_{22} \right )^{-1} A_{21} Y + \left ( sI -
  A_{22} \right )^{-1} B_2 U$$
Substituting into~(\ref{eq:LTIlaplace}) then yields
$$Y = W Y + V U$$
where $W=A_{11} + A_{12}\left ( sI - A_{22} \right )^{-1} A_{21}$ and
$V=A_{12}\left ( sI - A_{22} \right )^{-1} B_2 +B_1$.  Let $D$ be a
matrix with the diagonal term of $W$, i.e. $D=\mbox{diag}(W_{11}, W_{22}, ...,
W_{pp})$.  Then,
$$\left ( sI - D \right ) Y = \left ( W-D \right ) Y + V U$$
Note that $W-D$ is a matrix with zeros on its diagonal.  We then have
\begin{equation}
\label{eq:PQ}
Y =  QY + PU
\end{equation}
where 
\begin{equation}\label{eq:Q}
Q = \left ( sI - D \right )^{-1} \left ( W-D \right )
\end{equation}
and 
\begin{equation}\label{eq:P}
P=\left ( sI- D \right )^{-1} V   
\end{equation}
The matrix $Q$ is a matrix of transfer functions from $Y_i$ to $Y_j$,
$i \neq j$, or relating each measured signal to all {\it other}
measured signals (recall that $Q$ is zero on the diagonal).  The full
transfer matrix from $Y$ to $Y$ thus becomes $Q_T = I+Q$.  Likewise,
the transfer matrix from $U$ to $Y$ is $P$.

We thus can consider the transfer matrix, $N$, relating all manifest variables, $z$, to themselves.  This matrix is reminiscent of the structure matrix $S$ given in (\ref{eq:structure}), except that the entries are transfer functions relating variables instead of binary values.  this gives us the following definition.

\begin{definition}  Given the system (\ref{eq:LTI}), we define the {\it Dynamical Structure Function} or {\it Network}, $N$, of the system to be
\begin{equation}
\label{eq:fulldynamicstructure}
N = \left[\begin{array}{cc}I+Q&P\\0&I\end{array}\right].
\end{equation}
where $Q$ and $P$ are as given in~(\ref{eq:Q}) and~(\ref{eq:P}).
\end{definition}

When this function is completely characterized by $P$ and $Q$ (when
the system is open, that is, $u$ represents completely free inputs
unrelated to the measurements $y$), we refer to $(P,Q)$ as the {\it
  Dynamical Structure} of the system.  There are a number of
properties of the Dynamical Structure Function that makes it useful
for the structural analysis of linear systems:

\begin{proposition} Given the original realisation~(\ref{eq:LTI}),
  every entry $N_{ij}$ is a strictly proper function and unique.
\end{proposition}
Strict properness follows from the fact that $(sI-D)^{-1}$ (which is
strictly proper) is multiplying transfer functions that are at most
proper (never {\it improper}).  This fact is important for the
interpretation of $N$ as network {\it structure}.  The directed edges
associated with non-zero entries of this matrix imply {\it causal}
relations; strict properness of the transfer functions preserve this
interpretation.  Uniqueness follows by construction of both $Q$ and
$P$.

\begin{proposition} The transfer function, $G$, of the system (\ref{eq:LTI}), is related to Dynamic Structure by
\begin{equation}
G = \left(I-Q\right)^{-1}P.
\end{equation}
This fact follows directly from~(\ref{eq:PQ}) and $Y=GU$ and
demonstrates that Dynamic Structure is a factorization of a transfer
function into two parts, the {\it output} or {\it internal} structure,
$Q$, and the {\it input} or {\it control} structure, $P$.
\end{proposition}

It is now easy to see $N_{ij}=0$ if and only if there is no direct
{\it or hidden} connection from $z_j$ to $z_i$. The question is then
on how to determine the $p^2-p$ and $pm$ transfer functions in $Q$ and
$P$, respectively, to determine the Dynamical Structure from data.
This structure estimation, or reconstruction problem is addressed
next.

\section{Dynamical Structure Reconstruction}\label{se:main}

Assume data is collected from the original system~(\ref{eq:LTI})
leading to the transfer function in~(\ref{eq:G}) relating $Y = GU$.
Here we assume without loss of generality that $G$ is full rank.
Otherwise, there would be redundant inputs that could be removed to
get a full rank $G$.  Replacing $Y = GU$ in equation~(\ref{eq:PQ}) and
noting that the vector $U$ is abitrarely yields
\begin{equation}\label{eq:main}
(I-Q)G=P
\end{equation}

This equation shows that there are more unknowns than equations and
that in general Dynamical Structure of the $p$ measurable states
cannot be obtained from the $m$ inputs. There are $p^2-p$ unknowns in
$Q$, corresponding to all of the $Q_{ij}$ which represents the
internal Dynamical Structure. Then there are $pm$ unknowns in $P$
which represent the control Dynamical Structure on each measurable
state. Thus, all together, there are a total of $p^2-p+pm$ unknown but
only a total of $pm$ equations so the problem is under determined as
we have $p^2-p$ degrees of freedom. For instance, setting all
$Q_{ij}=0$ (which means no connection between measured states) and
$P=G$ is a solution of~(\ref{eq:main}) but probably the wrong one.

This clearly shows that the Dynamical Structure has {\em more}
information than $G$ and {\em less} than the original
system~(\ref{eq:LTI}).  Thus, to find the Dynamical Structure from $G$
we need {\em more} information. Either in the internal Dynamical
Structure (if we know some $Q_{ij}=0$), or on how the control is
affecting measurable state (if some $P_{ij}=0$).  Next we assume we
have no information on the internal Dynamical Structure (i.e. no
information on $Q$) and consider the cases where: $m<p$ (there are
less inputs than measured states), $m=p$ and $m>p$.  Before that, we
need the following technical result.

\begin{lemma}\label{lemma:rank}
  Rank$(P) =$ rank$(G)$.
\end{lemma}

\proof Since $(I-Q)G=P$, if suffices to show that rank$(I-Q)=p$.  It
follows that rank$(I-Q)$
$$= \ \ \mbox{rank} \left \{ I- \left ( sI - D \right )^{-1}\left (
  A_{11} + A_{12}\left ( sI - A_{22} \right )^{-1} A_{21} -D \right )
  \right \}$$
$$ \hspace{-8mm} =  \ \ \mbox{rank} \left \{ sI - D - \left ( A_{11} +
A_{12}\left ( sI - A_{22} \right )^{-1} A_{21} -D \right )\right \} $$
$$ \hspace{-20.5mm} = \ \ \mbox{rank} \left \{ sI - \left ( A_{11} +
  A_{12}\left ( sI - A_{22} \right )^{-1} A_{21} \right ) \right \}$$
which has rank $= p$. \myBox

\subsection{$m<p$: Less Inputs than Measured States}

If $m<p$, i.e. there are less inputs than measured states, and we have
no information on the internal Dynamical Structure then the Dynamical
Structure cannot be recovered.  To see this, note that in the best
case scenario $m=p-1$ and we would have $mp=p^2-p$ equations.  Since
there are $p^2-p$ unknowns from $Q$ we would need to know $P$
precisely.

The example from section~\ref{se:exam} shows how different networks
satisfy~(\ref{eq:main}) if $m<p$.  There we had two measurable states
$p=2$, a single input $m=1$ and $G=(G_{11},G_{21})$ given
by~(\ref{eq:examp_G}) .  In this case, equation~(\ref{eq:main}) has
two equations and four unknowns
\begin{equation}\label{eq:ex_21}
  \left\{\begin{array}{lll}
      G_{11} - Q_{12} G_{21} & = & P_{11} \\
      G_{21} - Q_{21} G_{11} & = & P_{21}
    \end{array}\right .
\end{equation}
We must solve for the internal Dynamical Structure ($Q_{12}$ and
$Q_{21}$) and the control Dynamical Structure ($P_{11}$ and $P_{21}$).
Since there are only two equations, there are two degrees of freedom.
A possible solution is to set $Q_{12} = Q_{21} = 0$, i.e. no internal
connection between $y_1$ and $y_2$.  In that case, $P_{11} =G_{11}$
and $P_{21} =G_{21}$ (Figure~\ref{fig:ex21a}). Note that $x_3$ is
playing the role of a hidden state (as $P_{21}$ is second-order) and
the system is not controllable (due to the common pole at $-1$), which
explains why $G$ is second-order and there are three states.  An
alternative is to have $P_{21}=0$ (which fixes $Q_{21}=G_{21}/G_{11}$)
and $Q_{12}=0$ (which fixes $P_{11}=G_{11}$), which can be seen in
Figure~\ref{fig:ex21b}.  Note that in this case $P_{11} \not =0$ as
that would result in a non-proper $Q_{12}$.  These two networks are
different and obviously only one can be correct.

\subsection{$m=p$: Same number of Inputs than Measured States}

\begin{theorem}\label{the:main}
  If $m=p$ and we have no information on the internal Dynamical
Structure, then the Dynamical Structure can be reconstructed if and
only if each input controls a measured state independently,
i.e. $P_{ij}=0$ for $i\not = j$.  In this case, the zeros of
$H=G^{-1}$ of the off diagonal define the internal Dynamical Structure
and
  $$Q_{ij} = -\frac{H_{ij}}{H_{ii}} \ \ \mbox{and} \ \ P_{ii} = \frac
  1 {H_{ii}}$$
\end{theorem}

\proof The ``if'' part of the proof follows from the fact that there
are $p^2 -p +p=p^2$ unknowns and $p^2$ equations. A linear set of
equations can be solved for $Q$ and $P$.  Multiplying on the right
$(I-Q)G=P$ by $H=G^{-1}$ yields $I-Q = PH$. Since $Q$ has zeros on its
diagonal and $P$ is diagonal, we have $1=P_{ii}H_{ii}$ or
$P_{ii}=1/H_{ii}$. Finally we can now solve for $Q=I-PH$ and the
result follows.

For the ``only if'', assume the Dynamical Structure can be
reconstructed, i.e. we can solve for $Q$ and $P$ in~(\ref{eq:main})
uniquely and they are all strictly proper.  Since rank$(G)=p$, by
lemma~\ref{lemma:rank} rank$(P)=p$. Thus, there are at least $p$
nonzero entries in $P$.

To show that there are at the most $p$ unknowns in $P$, assume there
are additional unknowns in $P$, and there is a Dynamical Structure
with strictly proper $Q^*$ and $P^*$ that satisfy~(\ref{eq:main}). We
want to show that another Dynamical Structure with strictly proper
$Q\not = Q^*$ and $P\not = P^*$ can be constructed.  Consider a vector
$X$ stacked with all the unknown parameters, i.e. with all unknown
$Q_{ij}$ and $P_{ij}$.  Equation~(\ref{eq:main}) can then be written
as ${\cal A} X ={\cal B}$, where both ${\cal A}$ and ${\cal B}$ are
functions of the elements of $G$.  Because there are $p^2$ equations
but $p^2+$ extra unknown elements in $P$, this system of equations is
undetermined or ${\cal A}$ has a null space.  Let $\bar X \not = 0$ be
an element of the null space of ${\cal A}$ and $X^*$ contain the
elements of $Q^*$ and $P^*$, which satisfy ${\cal A} X^* ={\cal B}$.
Then, there exists a large enough positive integer $n$ such that
$$X= X^* + \bar X \frac 1 {(s+1)^n}$$
is also a solution of ${\cal A}
X ={\cal B}$ and all the elements in $X$ are strictly causal.  We have
then found another Dynamical Structure which contradicts the
assumption.  Thus, at the most there are only $p$ unknowns in $P$.

Finally, there must then be exactly $p$ unknown and nonzero entries in
$P$. Since $P$ is full rank, each row and column must have exactly one
of these entries.  Without loss of generality the inputs can be
renamed and reordered so that the diagonal of $P$ contains the unknown
and nonzero entries.  \myBox

This theorem says that in addition to having a square and full rank
$G$ it is necessary and sufficient to know that each control $i$
affects first state $i$ before it affects any other measurable state
to reconstruct the Dynamical Structure. That allows to reduce the
number of unknowns to $p^2-p+p=p^2$ which can now be solved.

However, if there is some {\em a priori} information about the internal
Dynamical Structure (such as some of the $Q_{ij}=0$) then there is
more flexibility and less information and constraints are required of
$P_{ij}$.  As long as there are a total of $p^2$ nonzero elements
between $Q_{ij}$ and $P_{ij}$ then the Dynamical Structure can be
reconstructed by solving the linear system of
equations~(\ref{eq:main}).

If $P$ is not diagonal and we have additional information on how the
inputs affect the measured states, there may be a change of basis in
the control vector that allows it to be converted to a diagonal matrix
that can then be used in theorem~\ref{the:main}.  For example, if
$x_1$ is controlled by $u_1+u_2$ and $x_2$ by $u_1-u_2$ then one could
define two new input vectors $v_1=u_1+u_2$ and $v_2=u_1-u_2$.

If all the states are measured and $B=I$, we have the following
result.

\begin{corollary}\label{cor:full}
  If $p=m=n$ and $B=C=I$ then for $i\not = j$, $H_{ij}= a_{ij}$. Thus,
  $a_{ij}= 0$ ($i\not = j$) iff $H_{ij}=0$.
\end{corollary}

\proof The proof follows since is this case $G(s)= \left ( sI -A
\right )^{-1}$ which means $H(s)=sI-A$. \myBox

Note that we if knew we were measuring all the states we did not need
to know $B$. However, the information that the we measure all the
state is not captured by~(\ref{eq:main}), unless we had imposed that
there were only $n$ modes available to construct $Q$ and $P$.


\subsection{$m>p$: More Inputs than Measured States}

It may seem intuitive that if there are more inputs then there should
be more information.  However, the extra inputs are in fact
redundant. The reason is the fact that although $G$ is $p \times m$,
the rank$(G)=p$, which means that the inputs really only have $m$
degrees of freedom.  Thus, the problem reduces to having the same
number of inputs as measured states. The difference here is that we
may be able to choose from the $m$ inputs $p$ that are known to
control directly each measurable state.

\subsection{The Danger of Steady-State Measurements}

Before ending this section, we want to clarify some misconceptions,
especially some communities no so close mathematics, on steady-state
identification versus time-series data.  For instance,
in~\cite{gardner03} the authors proposed a method to estimate networks
based on full state measurement and control. From
corollary~\ref{cor:full}, the Dynamical Structure can be obtained from
the zeros of the non-diagonal terms of $H$, which correspond directly
to the entries $A$.  However, in the realistic case there are less
measurements and control available than states. If instead of
estimating $G$ from time-series data we were to use only steady-state
data, this could lead to mistakes as the following example shows.

Consider a third order system with measurements and control on the
first 2 states $x_1$ and $x_2$ and the following dynamics
$$A=\left[\begin{array}{ccc} -1 & 1 & -1 \\ 1 & -1 & -1 \\ 1 & 1 & -1
  \end{array}\right]$$
This is a fully connected network, and so we expect the reduced
network consisting on $x_1$ and $x_2$ to be fully connected as
well. In this case,
$$H(s)=\left[\begin{array}{cc}  \frac{s^2+2s+2}{s+1} & -\frac{s}{s+1} \\
          -\frac{s}{s+1} & \frac{s^2+2s+2}{s+1}
  \end{array}\right]$$
When $s \rightarrow 0$, 
$$H(0)=\left[\begin{array}{cc}  2 & 0 \\
          0 & 2
  \end{array}\right]$$
which could lead one to think the reduced order network is not
connected at all, i.e. $x_1$ does not affect $x_2$ and vice versa.  In
general, for third order systems this is always true if and only if
$a_{12}a_{33}+a_{13}a_{32}=0$ for the connection from $x_2$ to $x_1$
and $a_{21}a_{33}+a_{23}a_{31}=0$ for the connection from $x_1$ to
$x_2$. Note that even when these equalities are not exactly zero but
near zero, the presence of noise may again lead to wrong decisions.


\section{Conclusions}\label{se:conc}

This paper discussed the role of network structure for LTI systems.  In particular, it was shown that transfer functions alone contain no information about the structure of an LTI system.  We then introduced a new representation for such systems, a factorization of the system's transfer function that we call Dynamical Structure.  Dynamical Structure Functions contain more information about the system then the transfer function because they also describe the network structure between inputs and outputs.  Nevertheless, Dynamical Structure contains less information about the system than its state-space description because no attempt is made to realize the network structure relating the non-measured, hidden state variables to the rest of the system.  In this way, Dynamical Structure is a convenient analysis tool somewhere in between a system's full state space realization and its transfer function.

We then used Dynamical Structure to explore the network reconstruction problem.  In this problem, one would like to estimate network structure given only input-output data.  This problem is extremely important for a variety of fields, such as biology or counter-terrorism, that attempt to draw structural conclusions from data.  Necessary and sufficient conditions were presented that indicate that network reconstruction demands careful experiment design.  Moreover, various examples were provided throughout the paper that demonstrate how failure to respect the necessary conditions may lead to incorrect conclusions about the network structure.



\section{Acknowledgements} The authors would like to thank Glenn
Vinnicombe for his valuable input and comments, specially regarding
the notions of uniqueness and properness of $Q$ and $P$.

\bibliography{mybib}

\end{document}